# Vertex overload breakdown in evolving networks

Petter Holme[1,*] and Beom Jun Kim[2,1,†]

[1]*Department of Theoretical Physics, Umeå University, 901 87 Umeå, Sweden*
[2]*Department of Molecular Science and Technology, Ajou University, Suwon 442-749, Korea*

We study evolving networks based on the Barabási-Albert scale-free network model with vertices sensitive to overload breakdown. The load of a vertex is defined as the betweenness centrality of the vertex. Two cases of load limitation are considered, corresponding to that the average number of connections per vertex is increasing with the network's size ("extrinsic communication activity"), or that it is constant ("intrinsic communication activity"). Avalanche-like breakdowns for both load limitations are observed. In order to avoid such avalanches we argue that the capacity of the vertices has to grow with the size of the system. An interesting irregular dynamics of the formation of the giant component (for the intrinsic communication activity case is also studied). Implications on the growth of the Internet is discussed.

## I. INTRODUCTION

Natural and man-made systems for any kind of transportation or communication often take the form of large and sparse networks. Examples include neural networks [1], computer networks [2, 3], power grids [4], biochemical networks [5], and so on. Apart from their inherent randomness, these networks usually have some self-induced structure, which influences the flow of transport and robustness against congestion or breakdown in the network. A very common and conspicuous structure of many real-world networks is a power-law distribution of the degree (defined as the number of directly connected neighbors of a vertex), or scale-freeness for short [2, 3, 6, 7]. In practice, this means that a few vertices are much more central in the network flow than the others, and hence that the network is very sensitive to failure of these important vertices [2, 8, 9]. A limited capacity of the vertices can pose a serious limitation on the evolution of a growing network. This is especially serious for networks with an emerging scale-freeness, where the most loaded (and thus most important vertices) are far more loaded (important) than average.

Avalanche of breakdowns through the network is a serious threat when vertices are sensitive to overloading. A recent example is the blackout of 11 US states and two Canadian provinces on the 10th August 1996 [13]. This and similar examples serve as motivation for the study of how the extent and dynamics of the avalanches [10] are dependent on the network structure. Of particular interest is the overload breakdown problem in time evolving networks—as the network structure changes the load is redistributed, and if this is not accounted for it may trigger a vertex breaking avalanche. In the present paper we propose a model for breakdowns triggered by changing load in an evolving network. In this way the present work differs from earlier works on overload breakdown avalanches taking a fixed network as the starting point: Watts [11] modeled rare event cascades by letting individuals (vertices) change their behavior (a binary variable) if a threshold fraction of their neighbors have done that. Moreno, Gómez, and Pacheco investigated the fiber bundle model [12] on scale-free networks.

To investigate the time evolution of networks with an emerging power-law degree distribution sensitive to overload, we use the standard model for such networks—the Barabási-Albert (BA) model [14, 15], but with a maximum load capacity assigned to each vertex. The load is defined through the betweenness centrality a centrality measure for communication and transport flow in a network [16, 17]. If a vertex is overloaded, it breaks down by becoming disconnected from its neighbors.

## II. DEFINITIONS

In the following, we consider networks as graphs $G = (V, E)$ where $V$ is the set of vertices, and $E$ is the set of unweighted edges (unordered pairs of vertices). Multiple edges between the same pair of vertices are not allowed.

### A. Growth: Barabási-Albert model of scale-free networks

The standard model for evolving networks with an emerging power-law degree distribution is the Barabási-Albert model. In this model, starting from $m_0$ vertices and no edges, one vertex with $m$ edges is attached iteratively. The crucial ingredient is a biased selection of what vertex to attach to, the so called "preferential attachment": In the process of adding edges, the probability $P_u$ for a new vertex $v$ to be attached to $u$ is given

---

[*]Electronic address: holme@tp.umu.se
[†]Electronic address: beomjun@ajou.ac.kr

by [20]

$$P_u = \frac{k_u + 1}{\sum_{w \in V}(k_w + 1)},\quad (1)$$

where $k_u$ is the degree of the vertex $u$. We measure the time $t$ as the total number of added edges, which is different by factor $m$ from Refs. [14, 15] where $t$ is measured by the number of added vertices.

### B. Load and capacity

To assess the load on the vertices of a communication network, or any network where contact between two vertices are established through a path in the network, a common choice is the betweenness centrality [16]: Let $v$ be a vertex in $V$, then the betweenness centrality $C_B(v)$ is defined as

$$C_B(v) = \sum_{(w,w')} \frac{\sigma_{ww'}(v)}{\sigma_{ww'}},\quad (2)$$

where the summation is over all pairs of vertices such that $w \neq w'$ and $w, w' \neq v$, $\sigma_{ww'}$ is the number of geodesics (shortest paths) between $w$ and $w'$, and $\sigma_{ww'}(v)$ is the number of geodesics between $w$ and $w'$ that passes $v$. The betweenness centrality thus measures how many geodesics pass a certain vertex.

We below give a thorough motivation for the use of the betweenness centrality Eq. (2) as a load measure, and introduce two cases of assigning maximum load to each vertex. Suppose that $\Lambda$ is the set of pairs of vertices with established communications through shortest paths at a given instant [21]. Then let $\lambda(v)$ denote the *load* of $v$ defined as the number of geodesics that pass through $v$. If the pair $(w, w') \in \Lambda$ has many geodesics it contributes to $\lambda(v)$ with the fraction $\sigma_{ww'}(v)/\sigma_{ww'}$ of the geodesics that passes $v$. Since $\lambda(v)$ is of course highly dependent on the choice of $\Lambda$, we assume the effective load to be the average

$$\langle \lambda(v) \rangle_\Omega = \frac{1}{|\Omega|} \sum_{\Lambda \in \Omega} \lambda(v),\quad (3)$$

where $\Omega$ is an ensemble of $\Lambda$ (this is the assumption behind most usage of the betweenness centrality, although seldom stated explicitly). To proceed, we restrict $\Omega$ in two ways, one termed extrinsic communication activity (ECA) and the other one called intrinsic communication activity (ICA):

$$\Omega_{\text{ECA}} = \{\Lambda : |\Lambda| = A_{\text{ECA}}(N-1)(N-2)\},\quad (4a)$$
$$\Omega_{\text{ICA}} = \{\Lambda : |\Lambda| = A_{\text{ICA}}(N-1)\},\quad (4b)$$

where $A_{\text{ECA}}$ and $A_{\text{ICA}}$ are constants independent of $N$. This means that an element of $\Omega_{\text{ECA}}$ is a set of $A_{\text{ECA}}(N-1)(N-2)$ pairs of distinct vertices chosen uniformly at random (similarly for $\Omega_{\text{ICA}}$). In the first case of ECA, Eq. (4a), the average number of connections to a specific vertex is proportional to the system size $N$. In the second case of ICA, Eq. (4b), the average number of connections per vertex is constant. If we average $\lambda(v)$ over the respective ensembles we obtain

$$\langle \lambda(v) \rangle_{\Omega_{\text{ECA}}} = \frac{1}{|\Omega_{\text{ECA}}|} \sum_{\Lambda \in \Omega_{\text{ECA}}} \sum_{(w,w') \in \Lambda} \frac{\sigma_{ww'}(v)}{\sigma_{ww'}} =$$
$$= A_{\text{ECA}} C_B(v),\quad (5a)$$
$$\langle \lambda(v) \rangle_{\Omega_{\text{ICA}}} = \frac{1}{|\Omega_{\text{ICA}}|} \sum_{\Lambda \in \Omega_{\text{ICA}}} \sum_{(w,w') \in \Lambda} \frac{\sigma_{ww'}(v)}{\sigma_{ww'}} =$$
$$= \frac{A_{\text{ICA}}}{N-2} C_B(v) \approx \frac{A_{\text{ICA}}}{N} C_B(v).\quad (5b)$$

The results follow immediately from the definitions of the ensembles in Eq. (4): $\Omega_{\text{ECA}}$ is defined so that a pair of vertices $(w, w')$ occurs with a probability $A_{\text{ECA}}$; for $\Omega_{\text{ICA}}$ the same probability is $A_{\text{ICA}}/(N-2)$.

Now we introduce a capacity, or maximum value, $\lambda(v)^{\max}$ on the load that is the same for every vertex. From Eq. (5) we see that this is equivalent to having a constant maximum betweenness $C_B^{\max}$ in the ECA case, and to have a linearly growing $C_B^{\max}(N) = Nc_B^{\max}$ in the ICA case, where $c_B^{\max}$ is the parameter independent of $N$ [22]. If the load on a vertex $v$ exceeds this assigned value of maximum load, we consider the vertex $v$ overloaded, and all edges connecting to $v$ are deleted. However, the vertex $v$ is not removed and is still a member of $V$ and can thus be connected to in the future. After removing $v$'s edges, the load on other vertices might exceed the maximum load, so we recalculate $C_B$ of all vertices and remove edges of newly overloaded vertices. The above procedure is iterated until no vertices are overloaded.

To further motivate the study of the two cases ECA and ICA, we can consider the usage of a growing computer network. As the network becomes larger the amount of interesting information for a user becomes larger, so this mechanism contributes to an average increase of the others to communicate with activity per user, an increase proportional to $N$ corresponds to the ECA case. Communicating takes time, and assuming the time a user is willing to spend communicating restricts the number of communicators to a constant we have the ICA case. In reality both the above mechanisms are at work so one can expect the real behavior to be found between the ECA and ICA limits.

### C. Quantities for measuring network functionality

To measure the network functionality we consider three quantities, the number of edges $L$, inverse geodesic length $\ell^{-1}$, and the size of the largest connected subgraph $S$, all described in the following: For the original BA model the number of edges increases linearly as $L(t) = t$ (note that one edge is added in unit time). But if an overload breakdown occurs in the system $L$ decreases, making it a suitable simplest-possible-measure of the network functionality. In a functional network a large portion of the vertices should have the possibility to connect to each other. In percolation studies of random networks one often uses $S$ to define the system as 'percolated' (or functioning), when the size of the largest connected subgraph $S$ scales as $N$ [23]. One of the characteristic features of the BA model networks, as well as many real-world communication networks, is an logarithmically increasing average geodesic length $\ell$:

$$\ell \equiv \langle d(v,w) \rangle \equiv \frac{1}{N(N-1)} \sum_{v \in V} \sum_{w \in V \setminus \{v\}} d(v,w) , \qquad (6)$$

where $d(v,w)$ is the length of a geodesic between $v$ and $w$. As the average geodesic length is infinite when the network is disconnected (as could be the case when an overload breakdowns has occurred) we prefer studying the average inverse geodesic length [24]:

$$\ell^{-1} \equiv \left\langle \frac{1}{d(v,w)} \right\rangle \equiv \frac{1}{N(N-1)} \sum_{v \in V} \sum_{w \in V \setminus \{v\}} \frac{1}{d(v,w)} , \qquad (7)$$

which has a finite value even for the disconnected graph if one defines $1/d(v,w) = 0$ in the case that no path connects $v$ and $w$.

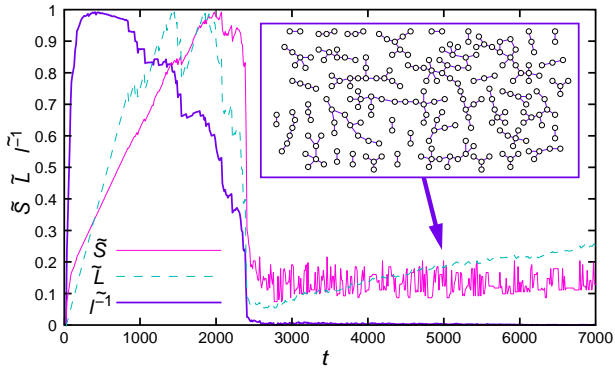

FIG. 1: The time evolution of $S$, $L$, and $\ell^{-1}$ for a typical run of the ECA case with $C_B^{\max} = 700$ and $m_0/2 = m = 12$. The curves display $S$, $L$, and $\ell^{-1}$ rescaled by their maximal values, $S_{\max} = 159$, $L_{\max} = 1133$, and $\ell_{\max}^{-1} = 0.2914$. In the inset the whole network at $t = 5000$ (except zero-degree vertices) is shown.

## III. EXTRINSIC COMMUNICATION ACTIVITY

A single run for the ECA load limit typically follows the time development shown in Fig. 1. The system grows exactly like the BA model until $C_B = C_B^{\max}$ for some vertex (this happens at $t \approx 840$ in Fig. 1). In this initial stage, the average inverse geodesic length approaches its maximum value very soon, and $S$ and $L$ grow linearly in time as expected. After reaching its early maximum ($t \approx 300$ in Fig. 1) $\ell^{-1}$ starts to decrease slowly since the average geodesic length increases in a connected growing network. As the network evolves, more vertices are overloaded but those breakdowns of overloaded vertices in this stage do not spread across the network but are soon recovered without causing severe damage to the whole network ($840 \lesssim t \lesssim 2200$ in Fig. 1). Finally, through a series of avalanches the system breaks down into small fragments of low average degree, characteristic for the large-$t$ steady state (this happens around $t \approx 2200$ in Fig. 1). Such a behavior has a natural explanation: In the emerging scale-free network (for $t \lesssim 1000$ in Fig. 1) the vertices with highest $C_B$ are also those with the highest degree [9]—when breakdown occurs the average degree decreases. But in a graph with fewer edges the geodesics must overlap to a larger extent, increasing the $C_B$ for less central vertices as well. As long as only few vertices are dominant the breakdown is limited to these top-end vertices. When the average degree is low enough (so the fraction of central vertices is high enough) the system looses the large connected component and will never recover. The large-$t$ equilibrium is characterized by many disconnected subgraphs of a low average degree. In a chain-like subgraph of size $s$ the maximal betweenness is of the order $\sim s^2/2$. Accordingly, when a new vertex joins two chain-like subgraphs of similar size the betweenness will increase by a factor around four, which in many cases is enough to cause a breakdown (in Fig. 1 the average time $\tau$ between breakdowns is $\tau \approx 5.8$ for $6000 \leq t \leq 7000$).

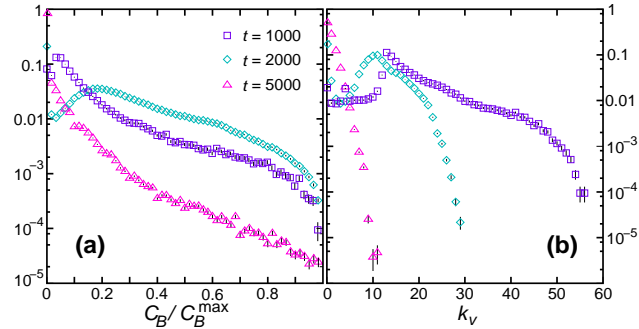

FIG. 2: Histograms for (a) the normalized betweenness $C_B/C_B^{\max}$ and (b) the degree $k_v$ for $C_B^{\max} = 700$ and $m = m_0/2 = 12$ at times $t = 1000$, $t = 2000$, and $t = 5000$.

From histograms of the betweenness and the average degree we can get another perspective on the breakdown. Figure 2 represents the same model parameters as Fig. 1, but the histograms are sampled over 500 independent runs. At the earliest time $t = 1000$—around the point when first vertex reaches the maximal load ($C_B^{\max} = 700$)—the betweenness distribution, as well as the degree distribution, shows an emerging power-law form. When vertices of high betweenness and degree breakdown (so the average degree gets lower), the number of vertices with an intermediate betweenness increases, as shown in Fig. 2(a) at $t = 2000$, close to the final fragmentation. After the final fragmentation has occurred ($t = 5000$ in Fig. 2(a)), the network approaches steady state and its structure does not change in time significantly, and the fraction of vertices with larger $C_B$ is reduced (compare $t = 2000$ and $t = 5000$ in Fig. 2(a)). In Fig. 2(b), the change of the degree distribution in time is displayed. With increasing $t$ the maximum degree decreases, but characteristic to all configurations before the final breakdown (at $t = 5000$ in Fig. 2(a)) is the existence of a peak in the histograms at an intermediate $k_v$.

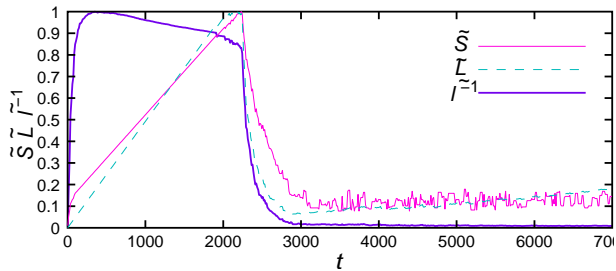

FIG. 3: A typical run of the BA model with random attachment instead of preferential attachment. The parameter values are those of Fig. 1. The curves display $S$, $L$, and $\ell^{-1}$, rescaled by their maximal values ($S_{\max} = 207$, $L_{\max} = 2030$, and $\ell^{-1}_{\max} = 0.2929$).

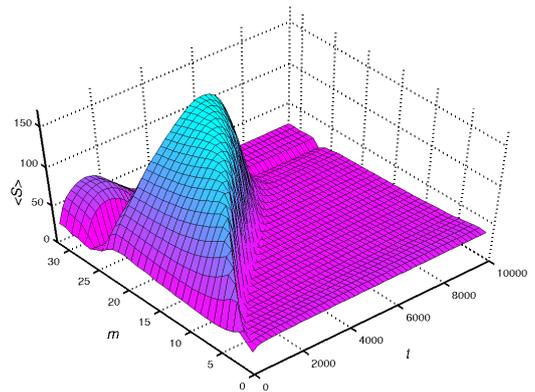

FIG. 4: The $m$ dependence of the time development of $\langle S \rangle$—the average size of the largest connected subgraph for the ECA case with $C_B^{\max} = 700$, and $m_0 = 2m$.

If the preferential attachment in the BA model's generative algorithm is replaced by a purely random selection of vertices to attach to (i.e., Eq. (1) is replaced by $P_u = 1/N$) the resultant network is known to have an exponentially tailed degree distribution [15]. Apart from probably being a relevant model for some evolving networks with an emerging exponential degree distribution, the randomly attaching BA model can serve as a background for studying the effects of preferential attachment.

Figure 3 shows a typical run with the same parameter values as those of Fig. 1 but with random instead of preferential attachment. Very large vertex breaking avalanches are known to occur in BA model networks under other load conditions (the fiber bundle model [12]). Thus it is no surprise that the network breaks down much faster when it is formed by preferential than random attachment. For example, the decay from $\tilde{S} = 0.8$ to $\tilde{S} = 0.2$ occurs in the the time interval $\Delta t \approx 550$ for random attachment in Fig. 3 whereas $\Delta t \approx 190$ is found for preferential attachment in Fig. 1. The relative robustness from random attachment is also consistent with the fact that Erdős-Rényi (ER) [25] model networks are more robust against vertex attack (most harmful removal of vertices) than the BA model networks [9].

The BA model with random attachment is not equivalent to the ER model in that all vertices of the former (except the initial $m_0$) have degree $k_v \geq m$. However, the vertices of arbitrary large importance present in the BA model with preferential attachment are—just as in the ER model—gone when the attachment is random. Thus, since both vertex attack and overload breakdown concern disabling the most important vertices, the similarity in behaviors is of little surprise. The relative robustness of the networks with random attachment is also reflected in the maximal values of $S$ and $L$: $S_{\max} = 159$ and $L_{\max} = 1133$ in Fig. 1, in comparisons with $S_{\max} = 207$ and $L_{\max} = 2030$ in Fig. 3. A more interesting feature of the BA model networks with random attachment is the lack of the second regime discussed in the context of Fig. 1, where the avalanche is limited and the network is able to recover. When the attachment is random the first avalanche almost always coincides with the maximum $S$ (and $L$). Before the first avalanche some vertices breaks without making other vertices overloaded. This ability to recuperate (for the case with the preferential attachment) can be interpreted as follows: When the most central vertex is overloaded and detached, the other vertices are not more loaded than that they can share the load of the additional geodesics. With random attachment, once the first vertex breaks, the others are too close in centrality to escape from being overloaded.

To investigate the $m$ dependence of $\langle S \rangle$ we average over more than 100 independent runs at various values of $m$ (the number of edge-additions for each new vertex). With an increasing average degree, the load is split over more vertices and thus the network becomes increasingly robust (this effect is dominant for $1 \leq m \leq 20$ in Fig. 4). But at the same time, a larger $m$ also means that a vertex, in the process of adding isolated vertices to a connected subgraph will get an increasingly high betweenness. These two competing effects cause the maximum of $\langle S \rangle$ at an intermediate $m$ ($m = 20$ in Fig. 4). The minimum of $\langle S \rangle$ near $m = 27$ in Fig. 4 can be explained as follows: When $C_B^{\max} = m(m-1)$, a new vertex becomes overloaded when it attaches its last edge, limiting $S$ to $m - 1$. In Fig. 4, this minimum occurs when $m = 27$ since $m(m-1) \approx C_B^{\max}(= 700)$.

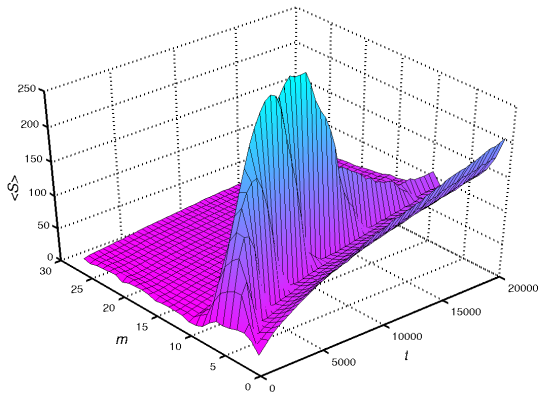

FIG. 5: The $m$ dependence of the time development of $\langle S \rangle$—the average size of the largest connected subgraph for the ICA case with $c_B^{\max} = 5$, and $m_0 = 2m$.

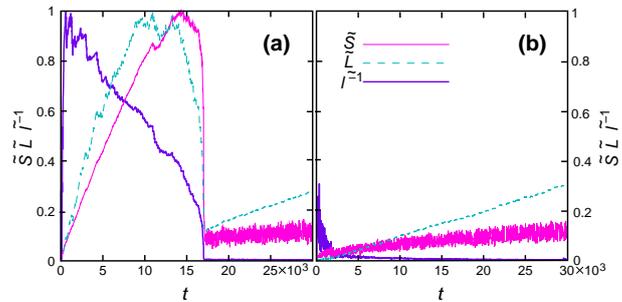

FIG. 6: Two runs with the ICA load limit and parameter values $m = m_0/2 = 12$ and $c_B^{\max} = 5$. In run (a) a giant component forms, in (b) no giant component emerges. The curves displays $S$, $L$, and $\ell^{-1}$ rescaled by their maximal values in (a) ($S_{\max} = 876$, $L_{\max} = 4354$, and $\ell_{\max}^{-1} = 0.2170$).

## IV. INTRINSIC COMMUNICATION ACTIVITY

Intrinsic communication activity (ICA), corresponding to the case the average number of connections per vertex is constant, is implemented through a linearly increasing maximum betweenness load $C_B^{\max}(t) = c_B^{\max} N(t)$. The $m$ dependence of the time evolution in the ICA case is shown in Fig. 5. Unlike the ECA case the maximum load is $m$ dependent ($C_B^{\max} = c_B^{\max} N = c_B^{\max}(m_0 + t/m)$), which causes the increase of $\langle S \rangle$ for small and decreasing $m$ in Fig. 5. Just as in the ECA case the networks grown with the ICA limit are most robust for an intermediate $m$ (with $c_B^{\max} = 5$ as in Fig. 5 the most robust networks are those with $m = m_{\max} \approx 10$). The mechanism behind this seemingly similar behavior is, however, entirely different from the ECA case. The avalanching breakdown follows the same pattern as the ECA case (shown in Fig. 1), but not the decrease in performance for $m > m_{\max}$: For some runs a giant component (a connected subgraph of the order of the whole system) fails to form, and the network stays disconnected, similarly to the large-$t$ equilibrium of the ECA case. The frequency of runs where no giant component is formed increases with $m$, causing the decrease of network functionality for $m > m_{\max}$. Two runs, one where a giant component is formed and one where it is not, are shown in Fig. 6. As seen in Fig. 6(a), the avalanche in the ICA case can be as dramatic as in the ECA case. This is expected since the relative increase of the maximum load, $\Delta C_B^{\max}/C_B^{\max}$, during the avalanche which lasts for $\Delta t$, is given by $\Delta C_B^{\max}/C_B^{\max} \approx \Delta t/t$, which becomes very small as $t$ increases. In other words, the increase of the maximum load during the avalanche is too slow to stop the global scale of breakdowns.

The existence of avalanche of breakdowns in the ECA case appears to be inevitable due to the fixed maximum load $C_B^{\max}$: As the network becomes larger more vertices are overloaded eventually, which leads to the situation that less edges should hold increased loads. In the ICA case, on the other hand, it is a very interesting question that why there has to be an avalanche at all, or, why a stable large connected component does not form. First we note that for a connected graph

$$\sum_{v \in V} C_B(v) = \sum_{w \in V \setminus \{v\}} (d(v, w) - 1) = N(N-1)(\ell - 1), \quad (8)$$

which yields

$$\langle C_B(v) \rangle = (N-1)(\ell - 1) \leq \max_{v \in V} C_B(v). \quad (9)$$

Consequently, Eq. (9) implies that the load of the vertex with the largest load in the network increases at least as fast as approximately $N\ell$ before the giant component becomes unstable. Since the BA network is well known to exhibit the small-world behavior that $\ell \propto \log N$ [14, 15], we conclude that in both ECA and ICA cases the giant component becomes unstable as $N$ becomes larger. The above argument suggests that to avoid the overload breakdowns the $C_B^{\max}$ should scale at least as $N^a$ with $a > 1$ and that if $a \leq 1$ (e.g., $a = 0$ for ECA and $a = 1$ for ICA cases in this paper) the giant component should break down as $N$ becomes larger. If we map this back to the original definition of load (see Sect. II B), we see that even in the ICA case the capacity $\lambda(v)^{\max}$ must grow (to be precise it has to grow at least as $N^{a-1}$, $a > 1$). If the growth of the network is exponential in time (as has been the case for the Internet [28]) the growth of the server capacities also has to be exponential. Fortunately, this has so far been the case.

Why does a giant component form only sometimes in the ICA case? Is it a long chain of events or just a singular attachment that establishes a giant component? To understand this we need some kind of criterion for a giant component to have formed. The basis of such a criterion is that $S$ scales like $N$ (we choose that $S$ should grow at least as $N/2$) over a certain interval. The problem is now to fix the interval: The lower limit should be quite high, so the system have time to form a giant component. At the same time the upper limit must be set before the avalanching breakdowns occurs, and this time decreases with decreasing $m$ (see Fig. 5). As a trade-off we say that a giant component has formed if

$$S > \frac{N}{2} \quad \text{in the interval} \quad 3m_0 \leq N(t) \leq 6m_0 ,\qquad (10)$$

where $N(t) = m_0 + t/m$ (and we choose $m_0 = 2m$). Now we can monitor the "basin of attraction" of the formation of a giant component in Fig. 6. Figure 7 shows the fraction $P_{\rm GC}$ of runs that forms a giant component when the network at time $t_{\rm start}$ equals that of Fig. 6 (a) and (b). From this we see that over a time of around $\Delta t = 50$ the formation of a giant component goes from a probability $P_{\rm GC} \approx 0.3$ to $P_{\rm GC} \approx 0.8$; a favorable sequence of attachments over $\Delta t = 50$ is thus needed to form a stable kernel, or starting point, for the formation of a giant component. The probability of a stable kernel to form decays fast with $t$ going from $\sim 20\%$ to $\sim 1\%$ during the initial 500 time steps. In Fig. 7 (b) we can see the formation of a stable kernel, that raises $P_{\rm GC}$ from 0.3 to 0.8, whereas Fig. 7 (c) shows the same time evolution in a case where no giant component forms. In the formation of the stable kernel we see many attachments to the connected component, and fewer attachments to isolated vertices. This attachment to the existent connected component works in favor of the establishment of a giant component in two ways: (1) The average degree of the connected component is increased since both more edges are included and less isolated vertices are joined. This makes more possible geodesics and enables more vertices to share the load. (2) The raise of average degree increases the probability of attachment to the connected component.

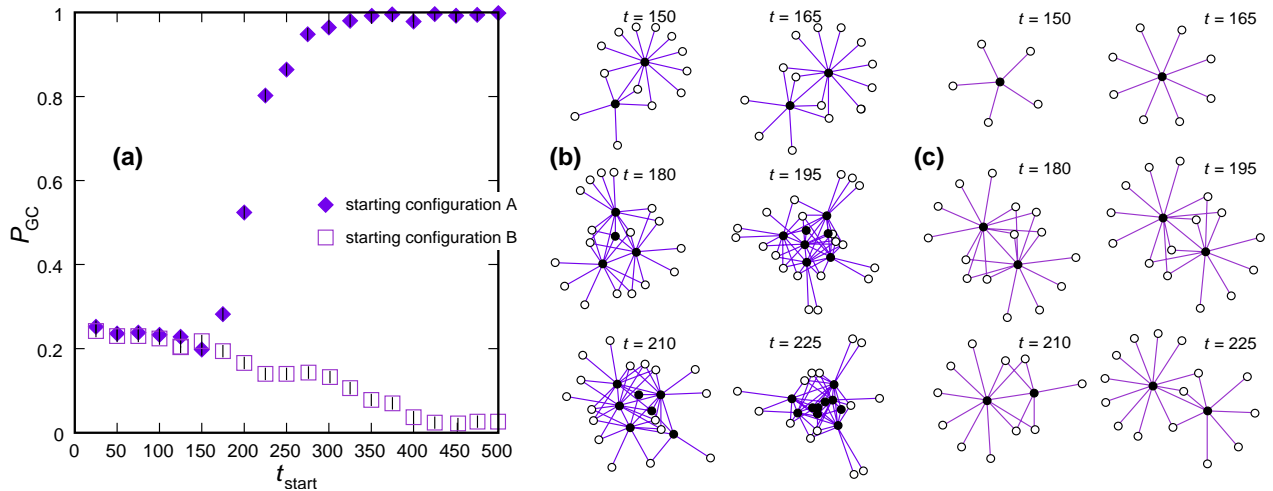

FIG. 7: (a) The fraction $P_{\rm GC}$ of runs that forms a giant component when the network at time $t_{\rm start}$ equals that of Fig. 6 (a) and (b). A giant component is said to have formed if the condition (10) is fulfilled, i.e., $S > N/2$ when $864 \leq t \leq 1728$. (b) The network of Fig. 6 (a) during the formation of a stable kernel $175 \leq t \leq 225$ of many vertices sharing the load. (c) The network of Fig. 6 (b) during the same time. Vertices of nonzero betweenness are indicated by filled circles. Vertices of zero degree are not displayed. Graph drawing was aided by the Pajek package [26].

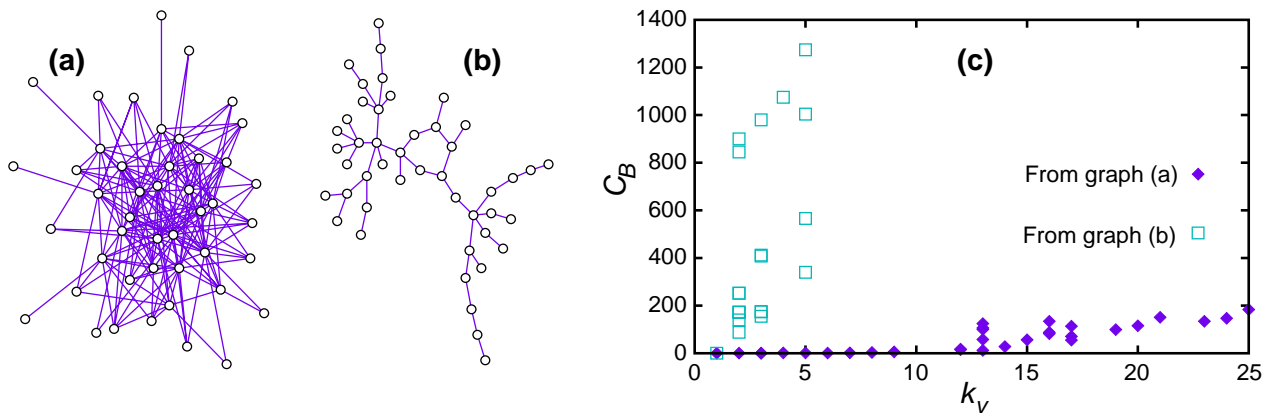

FIG. 8: (a) The largest connected subgraph at $t = 385$ in Fig. 6 (a). (b) The largest connected subgraph at $t = 10000$ in Fig. 6 (b). (c) shows scatter plot of betweenness vs degree in the graphs displayed in (a) and (b).

The reason why we do not observe any emerging gi-ant components in the large-$t$ limit is evident from Fig. 8. In the large chain-like subgraphs (see Fig. 8(b)) dominant in the large-$t$ limit, there are many vertices of high betweenness. Joining two such graphs gives a large increase in betweenness to most vertices, as opposed to Fig. 8(a) where an additional load can be shared by many vertices, and the much shorter average geodesic length leads to a much lower total increase of $C_B$. A stable kernel is of course not forbidden to form, but the probability of formation decreases with time (as seen in Fig. 7(a)). Whether or not the probability of a stable kernel to form as $t \to \infty$ is less than one is an interesting open question.

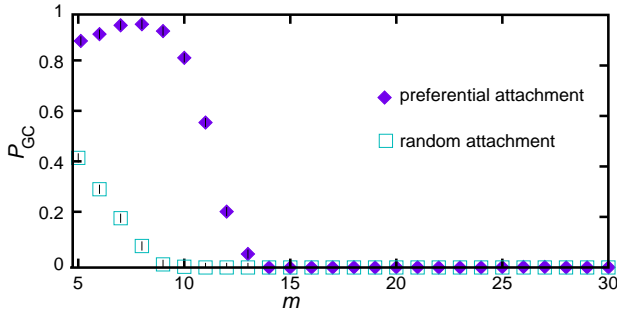

FIG. 9: The fraction $P_{\text{GC}}$ of runs that forms a giant component for preferential and random attachment as a function of $m$. The decay of the preferential-attachment curve as $m \to 0$ is due to early avalanches (the increase of such for decreasing $m$ makes the rule of thumb criterion Eq. 10 inapplicable for $m < 5$, thus these are not shown).

To proceed we investigate the effect of preferential attachment for the formation of a giant component. For the random attachment case giant components form very seldom, the reason is that the stable kernels are more likely to form with preferential attachment, since then attachment are done mainly to highly connected areas which are likely starting points for a stable kernel. Figure 9 shows the probability of formation of a giant component (by the condition Eq. (10)) for preferential and random attachment. For growing $m$ and random attachment the probability of giant component formation $P_{\text{GC}}$ is strictly decreasing. This can be understood since more and more isolated vertices become connected as $m$ increases, thus leading to the time development seen in Fig. 7 (c) and discussed above. For preferential attachment the decrease occurs at $m \gtrsim 8$. In the large $m$ limit, the probability of the momentary higher-than average increase of the average degree needed for the formation of a stable kernel decreases; thus the decrease for large $m$.

## V. SUMMARY AND CONCLUSIONS

We have used the BA model with the preferential attachment and investigated the development of networks where the vertices might break down due to overload. The load was defined in terms of the betweenness centrality by careful arguments. We considered two cases of load limitation: One corresponding to that the average number of connections per vertex is increasing with the number of vertices of the network (the ECA case), and one where the average number of connections per vertex is constant (the ICA case).

We find that for both the ECA and ICA cases overloading may cause breakdown avalanches. At a critical point these avalanches will fragment the network completely and it will never recover. The mechanism is that the vertices with the highest loads are also the vertices with the highest degrees, and accordingly removing such vertices will thus decrease the average degree mostly and increase the betweenness on other vertices maximally. In the large time equilibrium we find that the network consists of many isolated chain-like clusters.

Even in the seemingly less restricted ICA case we find that the network will eventually become fragmented, and thus that for the network to be connected the capacity of the vertices to relay connections have to increase with the size of the network. Even if the congestion of computer network traffic will result in a slowing down of the servers, rather than a complete breakdown, we anticipate that this will be a problem in the Internet if the exponential increase of computer performance stalls but not the growth of the number of Internet sites. Overflow control is implemented in many communication networks, such as telecommunication networks [27] and to some degree in large-scale Internet routing protocols such as the Border Gateway Protocol [28]. We suppose that these kind of overload control strategies can be optimized, and centers of potential overload be removed by a careful study of the network geometry.

In the ICA case we find that a giant component not necessarily forms every run, and that a particularly stable configuration has to be formed during a short period of time to trigger the emergence of a giant component. Furthermore, preferential attachment greatly increases the probability of a giant component to form compared to random connection. This might be a further explanation to preferential attachment in natural systems such as the metabolic networks discussed in Ref. [29].

### Acknowledgements

The authors acknowledges support from the Swedish Natural Research Council through Contract No. F 5102-659/2001.